\documentclass[fleqn,twoside]{article}
\usepackage{gc}
\usepackage{amsfonts,amsmath,amssymb,amscd,euscript}

\heads{Vladimir V. Kassandrov}
      {}

\newcommand{\be}{\begin{equation}\label}
\newcommand{\ee}{\end{equation}}
\newcommand{\prt}{\partial}
\newcommand{\p}{\prime}
\newcommand{\bib}{\bibitem}
         
\newcommand{\wh}{\widehat}

\begin{document}
\twocolumn[

\Title{ELECTROMAGNETIC FIELDS       \yy
       IN KERR-SHILD SPACE-TIMES}


   \Author{Vladimir V. Kassandrov            
              }   
         {Institute of Gravitation and Cosmology, Peoples' Friendship University of Russia, 6 Mikluho-Maklay St., \\ Moscow 117198, Russia}              


\Abstract
{Making use of twistor structures and the Kerr theorem for shear-free 
null geodesic congruences, an infinite family of electromagnetic fields 
satisfying the homogeneous Maxwell equations in flat Minkowski and 
the associated curved Kerr-Schild backgrounds is obtained for any such 
congruence in a purely algebraic way. Simple examples of invariant 
axisymmetric Maxwell fields are presented.

}


]  

\section{Introduction}

Rather unexpectedly, one observes a growing interest in solutions of the 
most investigated, seemingly simple and important equations of classical field 
theory, the  Maxwell linear homogeneous  equations in vacuum. It occures that 
these possess various kinds of solutions with different topology of 
field lines~\cite{Ranada,Keihn}, with extended singularities of complicated shape and 
temporal dynamics~\cite{Joseph,Trishin,Sing}, multi-valued complexified solutions etc. 
The simplest example of the latter is provided by the electromagnetic field of the  
Kerr-Newman solution in general relativity (which is quite the same 
in the curved Kerr-Newman background and in flat Minkowski space). 

To obtain complicated solutions to the Maxwell equations, various 
algebraic methods can be used, which involve twistor structures or 
generating procedures starting from a solution to some (linear or nonlinear) 
``master equation'' (the concepts of a superpotential or ``hidden nonlinearity'', 
respectively). In the later case, one obtains natural restrictions, a sort of 
``selection rules'', on characteristics of the associated Maxwell fields, 
in particular, on admissible values of the electric charge of isolated 
singularities~\cite{Ranada,Sing,GR95}.

It is especially noteworthy that, in many cases, solutions to the Maxwell 
equations in flat space-time $\eta_{\mu\nu}$ are exactly invariant under a 
deformation of the metric to the following {\it Kerr-Shild form}:
\be{KerrShild}
g_{\mu\nu} = \eta_{\mu\nu} + H(x) k_\mu k_\nu , 
\ee
$k_\mu (x)$ being a null 4-vector field. Indeed, 
for a determinant of any metric of the form (\ref{KerrShild}) one obtains 
$\sqrt{-g}\equiv 1$, whereas the condition
\be{raise}
g^{\mu\rho}g^{\nu\lambda} F_{\rho\lambda} = \eta^{\mu\rho}\eta^{\nu\lambda} F_{\rho\lambda}:= F^{\mu\nu}
\ee
is fulfilled iff the 4-vector $k_\mu$ is an eigenvector of the field strength 
tensor $F_{\mu\nu} = \prt_\mu A_\nu - \prt_\nu A_\mu$, that is, iff 
\be{eigen}
F_{\mu\nu} k^\nu \sim k_\mu .
\ee
The well known examples of such an invariance of Maxwell field under a 
Kerr-Shild deformation of the flat metric are the ordinary Coulomb field in 
the Reissner-Nordstr\"om background or the Kerr-Newman field mentioned above. 

In Section 2, we describe a procedure which makes use of 
twistor structures and allows us to associate a whole family of  
electromagnetic fields {\it with any shear-free null congruence (SFNC)} of rays 
in Minkowski space. We demonstrate that the condition (\ref{eigen}) {\it holds  
identically for all such fields} if $k_\mu$ is a null 4-vector field tangent 
to the SFNC under consideration. Consequently, any SFNC gives rise to a 
family of Kerr-Shild metrics (with different $H(x)$ in (\ref{KerrShild})) and 
a family of electromagnetic fields identically satisfying Maxwell equations 
\be{Maxwell}
\prt_\nu (\sqrt{-g} F^{\mu\nu}) = 0
\ee
in the corresponding curved background. In addition, it is well known that 
the scalar ``gravitational potential'' $H(x)$  may be 
often fitted to ensure that the Kerr-Schild metric satisfies the vacuum or 
electrovacuum Einstein equations. 
  
In Section 3, we present simple examples of invariant electromagnetic fields 
associated with SFNC and inheriting the spherical or axial symmetry of the 
congruence. In all such cases the newly found fields correspond to the 
previously determined ones (which are known to possess the invariance property). 
In general, however, the invariant fields introduced in the paper are new and 
can open a way to complicated electrovacuum solutions with metrics of the 
Kerr-Schild type.

\section{Shear-free null congruences and the associated Maxwell fields 
in flat and curved background} 

According to the {\it Kerr theorem}~\cite{Penrose}, any SFNC in Minkowski 
space may be obtained from the generating equation
\be{Surface}
\Pi(G, \tau^1, \tau^2) \equiv \Pi(G, wG+u, vG+\bar w)= 0, 
\ee     
where $\Pi$ is an analytical function of three complex arguments 
$\{G, \tau^1, \tau^2\}$ representing the components of a {\it projective 
twistor}. The latter is associated with the coordinates $u,v = t \pm z,~ 
w,\bar w = x \mp Iy$ of  
points in Minkowski space through the Penrose's {\it incidence relation} 
\be{inc}
\tau^1 = wG+u,~~~ \tau^2 =  vG+\bar w.
\ee
Resolving (\ref{Surface}) with respect to the unknown G, one comes to a 
multi-valued field $G = G(u,v,w,\bar w)$. It is easy to demonstrate 
(via differentiaton of (\ref{Surface})) that any continious branch of this 
multi-valued field satisfies the determining equations of a SFNC, 
\be{SFNC}
\prt_w G = G\prt_u G, ~~~\prt_v G = G \prt_{\bar w} G, 
\ee
and therefore represents the principal spinor of the latter. Thus, one has 
a one-to-one correspondence between the SFNCs and the twistor functions 
$\Pi(G,\tau^1,\tau^2)$ (precisely, the surfaces $\Pi=0$ in the ${\mathbb C}P^3$ 
projective twistor space). 

From (\ref{SFNC}) the complex eikonal equation follows immediately, 
\be{eik}
| \nabla G |^2: = \prt_u G \prt_v G - \prt_w G \prt_{\bar w} G  = 0, 
\ee
and, as the integrability condition of (\ref{SFNC}), one obtains then the 
d'Alembert equation~\cite{Asya} 
\be{dAlambert}
\Box G: = \prt_u \prt_v G - \prt_w \prt_{\bar w} G = 0.  
\ee

One might thus suspect that any SFNC gives rise to a (complexified) 
Maxwell field, with $G$ being a sort of super-potential. Indeed, 
in~\cite{Joseph,GR95,Asya} such a field has been obtained, with its strength 
expressed through the second-order derivatives of $G$. Electromagnetic 
fields of this type possess a number of the afore-mentioned peculiar properties.  
However, these fields do not preserve their form under the Kerr-Schild 
deformation of metric. To ensure such an invariance, we propose below another  
expression for Maxwell fields to be associated with arbitrary SFNC. 

Specifically, consider the following simple ansatz for the components of the 
spintensor $\varphi_{A^\p B^\p},~A^\p,B^\p =1^\p,2^\p$ of  
electromagnetic field strength:
\be{spcomp}
\varphi_{1^\p1^\p} = \prt_u\prt_u F,~\varphi_{1^\p2^\p} = \varphi_{2^\p1^\p} = 
\prt_w\prt_u F,~\varphi_{2^\p2^\p} = \prt_w\prt_w F, 
\ee
where $F=F(G)$ is an arbitrary (holomorphic) function of the principal 
spinor component $G(X)$.

Taking into account that $X=\{X^{AA^\p}\} = \{u,w,{\bar w},v\}$, it is easy to 
check that the field (\ref{spcomp}) in fact satisfies the homogeneous Maxwell 
equations 
\be{spMaxwell}
\prt^{AA^\p} \varphi_{A^\p B^\p} = 0 
\ee
provided the eikonal (\ref{eik}) and d'Alembert (\ref{dAlambert}) equations 
hold both together for the function $G$.

Now, let us prove that all fields (\ref{spcomp}) associated with a SFNC 
always obey the eigenvector condition (\ref{eigen}). To do so, let us  
rewrite the latter in the equivalent 2-spinor form
\be{speigen}
\varphi_{A^\p B^\p} \xi^{A^\p} \xi^{B^\p} = 0, 
\ee
where $\xi_{A^\p}$ is the principal spinor of the SFNC in the gauge 
$\xi_{1^\p} = 1,~ \xi_{2^\p} = G$. Substituting the components (\ref{spcomp}) 
into the above condition (\ref{speigen}) and transforming the l.h.s. of the 
latter, one obtains
\be{transform}
G\prt_u (F^\p M) - \prt_w (F^\p M) - (F^\p M) \prt_u G = 0, 
\ee
where $F^\p:=dF(G)/dG,~~M:= G\prt_u G - \prt_w G$.  However, $M\equiv 0$ on account of 
the first of the SFNC determining equations (\ref{SFNC}), so that the eigenvector 
condition (\ref{speigen}) is identically satisfied for the fields 
(\ref{spcomp}). This means that 
{\it there exists an infinite family of electromagnetic fields (\ref{spcomp}) 
associated with a SFNC of a general form, which all satisfy the homogeneous Maxwell 
equations in Minkowski space and in the corresponding curved 
Kerr-Schild space}. 

Some remarks are in order here. 

1. In a number of papers E.T. Newman et al. and A.Ya. Burinskii (see, e.g.,
~\cite{Lind,Burin}) had been working with electromagnetic fields generated by 
a pointlike charge moving in real or complexified Minkowski space. These  
fields indeed satisfy the invariance condition (\ref{eigen}). However, even for 
this particular case,  to obtain such fields, explicit integration of 
the Maxwell equations was required.     

2. To simplify the exposition, nearly all the above constructions 
have been presented in a particular gauge. One understands, nonetheless, that 
the formalism can be easily transformed into a manifestly Lorentz invariant 
form.

3. For the newly found fields, the theorem on ``quantization of the electric 
charge'' for isolated singularities~\cite{Sing} remains valid and will be reproduced 
elsewhere (see also the examples below). 

4. Remarkably, symmetries of the electromagnetic fields (\ref{spcomp}) for 
a general form of $F(G)$ can be weaker than those of the SFNC which they are 
generated from. For example, the Kerr axisymmetric congruence can give rise 
to electromagnetic fields devoid of any symmetries at all. To preserve the 
axial symetry of a SFNC, a particular form of the function $F(G)$, namely 
$F(G)=G^{-1}$, must be used (see below).

\section{Some examples of invariant Maxwell fields associated with a SFNC}

1. Consider first a static, spherically symmetric SFNC and the associated fields 
corresponding to the following form of generating equation (\ref{Surface}):
\be{Coulomb}
\Pi=G\tau^1 - \tau^2 \equiv w G^2  -2zG +{\bar w} =0, ~~z= (u-v)/2. 
\ee
Resolving (\ref{Coulomb}), one obtains 
\be{Stereo}
G= \frac{\bar w}{ z\pm r}, ~r:= \sqrt{w{\bar w}+z^2}\equiv \sqrt{x^2+y^2+z^2}, 
\ee   
i.e. {\it the stereographic projection} $S^2\mapsto {\mathbb C}$ from the South 
or North poles, respectively. Corresponding SFNC is the radial, spherically
symmetric congruence with a point singularity.  

One can easily verify that, from the whole family of associated 
electromagnetic fields (\ref{spcomp}), only the choice $F(G)= G^{-1}$ 
corresponds to a field inheriting the spherical symmetry of the congruence.   
Specifically, calculating the components (\ref{spcomp}) of the 
spintensor $\varphi_{A^\p B^\p}$ for $F(G)= G^{-1}$ , one gets 
\be{Coulfield}
\varphi_{1^\p 1^\p}= q \frac{w}{r^3},~\varphi_{1^\p 2^\p}= -q \frac{z}{r^3},
~\varphi_{2^\p 2^\p}=  -q \frac{\bar w}{r^3}, 
\ee
with the constant $q=\mp 1/4$. The expression (\ref{Coulfield}) describes the  
pure Coulomb field with an electric charge necessarily 
fixed in value (in fact this is a minimum ``elementary'' charge, 
see~\cite{Sing,GR95} for details). Of course, the Coulomb field is invariant 
under the Kerr-Schild type of deformation of flat metric, i.e., under  
transition to the Reissner-Nordstr\"om space-time. 
      
2. The {\it Kerr congruence} with twist and a ring-like singularity of radius 
$a$ is known to arise from the radial one through a complex shift, say, 
$z\mapsto z +Ia$. Since all the above-presented constructions deal with 
complex holomorphic structures, the associated Maxwell field also 
follows from (\ref{Coulfield}) through such a shift and 
coincides with the electromagnetic field of the Kerr-Newman electrovacuum  
solution. It is important, nonetheless, that in the same Kerr-Newman 
metric background, making use of different $F(G)$, a lot of other fields   
obeying the Maxwell equations could be introduced. 

3. The Kerr congruence can be naturally generalized to a {\it nonstationary} 
form  described in~\cite{GR09}. To do that, one transforms the principal spinor   
of the radial SFNC (\ref{Stereo}) through a {\it complex boost} and 
obtains the following form of it: 
\be{defspinor}
G=(1+ Iu) \frac{x+Iy}{(z-z_t) \pm \wh r},~~z_t:=-Ia+Iut,
\ee
where $Iu\in {\mathbb C}$ is the ``imaginary velocity'' parameter of a ``complex boost'', 
and ``complex distance'' from the point singularity to an observation point is
\be{cdist}
\wh r = \sqrt{(z-z_t)^2+w{\bar w}(1+u^2)}.  
\ee
The singular locus (caustic) of the SFNC determined by (\ref{defspinor}) coincides 
with the {\it branching points} $\wh r=0$ of the principal spinor and represents 
 a {\it Kerr-like ring} collapsing/expanding with the physical velocity $V=u/\sqrt{1+u^2} <1$. 
The invariant (complex-valued) electromagnetic field associated with 
such a congruence through the expression (\ref{spcomp}) 
under the assumption $F(G)=G^{-1}$ again 
reproduces the previously determined one (see Eq.(40) of the paper~\cite
{GR09}) and manifests a number of interesting properties (including its  
vortex-like structure and concentration of the field along the symmetry axis).    

4. We thus see that, for the simplest SFNCs, the invariant electromagnetic 
fields (\ref{spcomp}) contain a distinguished one, 
inheriting the axial symmetry of the congruence and reproducing the fields  
associated with the latter through the old prescription. Generally, 
however (in particular, for the case of a {\it bisingular SFNC}~\cite{Joseph,Asya}), 
the newly obtained (and necessarily invariant) Maxwell fields differ  
from the previously considered (and generally noninvariant) ones. 
One can hope thus that discovery of a family of fields automatically 
satisfying the Maxwell equations in any given Kerr-Schild background will 
open the way to construction of extremely complicated 
electrovacuum solutions corresponding to arbitrary SFNCs.

\end{document}